\documentclass[11pt]{article} 
 
\usepackage{amssymb,cite} 
\usepackage{graphicx} 
\usepackage{latexsym} 
\usepackage{epsfig,epic,amsfonts} 
\usepackage{wrapfig} 
\def\thefootnote{\fnsymbol{footnote}} 
 
\newcommand{\eq}{\begin{equation}} 
\newcommand{\en}{\end{equation}} 
\newcommand{\be}{\begin{equation}} 
\newcommand{\ee}{\end{equation}} 
\newcommand{\eqa}{\begin{eqnarray}} 
\newcommand{\ena}{\end{eqnarray}} 
\newcommand{\ba}{\begin{eqnarray}} 
\newcommand{\ea}{\end{eqnarray}}

\newcommand{\bra}{\left \langle}
\newcommand{\ket}{\right \rangle}

\newcommand{\ZZ}{\hbox{{\rm Z{\hbox to 3pt{\hss\rm Z}}}}}

\newcommand{\EQ}{\begin{equation}} 
\newcommand{\EN}{\end{equation}} 
\newcommand{\bea}{\begin{eqnarray}} 
\newcommand{\eea}{\end{eqnarray}}

\begin{document} 
\begin{titlepage} 
\vskip0.5cm 
\begin{flushright} 
\end{flushright} 
\vskip0.5cm 
\begin{center} 
{\Large\bf On the intrinsic width of the chromoelectric flux tube in finite temperature LGTs.} 
\end{center} 
\vskip1.3cm 
 
\centerline{    
M. Caselle$^{a}$ and P. Grinza$^{b}$}    
    
 \vskip0.4cm    
 \centerline{\sl  $^a$ Dipartimento di Fisica    
 Teorica dell'Universit\`a di Torino and I.N.F.N.,}    
 \centerline{\sl via P.Giuria 1, I-10125 Torino, Italy}    
 \centerline{\sl    
e--mail: \hskip 1cm    
 caselle@to.infn.it}    
 \vskip0.1 cm    
 \centerline{\sl Departamento de Fis\'ica de Part\'iculas,}
 \centerline{\sl Universidad de Santiago de Compostela, 15782 Santiago de
Compostela, Spain}    
 \centerline{\sl    
e--mail: \hskip 1cm grinza@fpaxp1.usc.es}    
 
 \vskip0.4cm    
    
\begin{abstract}    
We propose three different lattice operators to measure the intrinsic width $\xi_I$ of the chromoelectric flux tube in pure lattice gauge theories. 
In order to test these proposals we evaluate
them for SU(2) and Ising LGTs in (2+1) dimensions in the vicinity of the deconfinement transition. Using dimensional reduction, we could perform the calculation 
in the effective 2d spin model using standard S-matrix techniques. We consistently found the same result for the three lattice operators. This result
can be expressed in terms of the finite temperature string tension as follows $\xi_I=\frac{T}{2\sigma(T)}$ and implies
that the intrinsic width of the flux tube diverges as the deconfinement transition is approached.
\end{abstract}    
\end{titlepage}    
 
\setcounter{footnote}{0} 
\def\thefootnote{\arabic{footnote}} 
 
\section{Introduction} 
 In these last years much interest has been attracted by the study of the flux tube thickness in the confining regime of Lattice Gauge Theories (LGTs).
At low temperatures the square width of the flux tube is predicted to increase logarithmically with the interquark distance~\cite{lmw80}. This prediction has been confirmed
by numerical simulations in various pure gauge theories, first in abelian models ~\cite{cgmv95, Zach:1997yz, Koma:2003gi, Panero:2005iu, Giudice:2006hw} and more recently 
also in non-abelian 
LGTs~\cite{Gliozzi:2010zv,Bakry:2010zt} (see also ~\cite{Bali:1994de} for some early attempt). 

The situation changes drastically as the deconfinement temperature is approached from below. In fact it can be shown that in this regime the dependence of the square width
on the interquark distance becomes linear with a proportionality constant which diverges as the deconfinement transition is approached\cite{Allais:2008bk,Caselle:2010zs}.
 Also these predictions were
nicely confirmed by numerical simulations\cite{Caselle:2010zs,Gliozzi:2010jh}.

All the theoretical predictions mentioned above were obtained using as effective string theory the Nambu-Goto action. 
However we know that the Nambu Goto action is consistent at the quantum level only in 26 dimensions and as a consequence 
for any LGT in three or four dimensions the correct effective string action, whatever it is, should deviate
from the Nambu-Goto one at some high enough order in the perturbative expansion. The consequences of these deviations can be seen, for instance, looking at the 
critical behaviour of the theory in the vicinity of the deconfinement transition. For those models which undergo a second order deconfinement transition
the Nambu Goto action predicts mean field critical indices while the Svetitsky-Yaffe
analysis~\cite{sy82} (confirmed by a host of numerical simulations) predicts non trivial universality classes for the different LGTs, depending on 
the center of the gauge group.

These deviations from the Nambu-Goto action may be of two independent types: they can be due to irrelevant or marginal terms in the effective string action or they can be 
due to the coupling of the massless degrees of freedom of the effective string with massive (non-stringy) "intrinsic" excitations of the flux tube. Due to our incomplete 
understanding of how the effective string description emerges from QCD we have no precise description of these massive modes, of their dynamical origin 
and of their action, but we have a few hints which may help our intuition. The most important one is that these massive modes 
should manifest themselves as a sort of 
"{\sl intrinsic width}" of the flux tube. The Nambu Goto action in fact describes a string of vanishing
intrinsic width and the flux tube thickness discussed in~\cite{lmw80}  (which we shall call in the following "{\sl effective width}" 
to avoid confusion) is entirely due to the quantum fluctuations. 
In this respect the intrinsic width, which we shall denote in the following as $\xi_I$, 
can be viewed as the residual thickness of the flux tube when the
interquark distance $R$ is pushed down to the scale (typically $R\sim 1/\sqrt{\sigma_0})$ below which the
effective string description does not hold any more and does not contribute to the flux tube thickness.

We expect that the intrinsic width should not depend on the
interquark distance $R$ (in opposition to the effective width which instead increases with $R$) but should instead be a new fundamental scale of the theory.
It is tempting to identify this intrinsic width with the size of the Nielsen-Olesen vortex line (or equivalently the London penetration length of the dual
Abrikosov vortices\footnote{See~\cite{Cardaci:2010tb} for un updated discussion of this issue.})
in the framework of the dual superconductor models of confinement which however has been proved rigorously 
only for Abelian Higgs or Georgi-Glashow like models. 
We shall further comment on this issue in the following. Let us also mention among the attempts to characterize
the intrinsic width the holographic model recently discussed in \cite{Vyas:2010wg}.

Thanks to the remarkable universality theorems
proved in~\cite{lw04,A2009} we know that the first few perturbative orders (in the expansion in powers of $1/\sigma_0 R^2$ of effective string action) are universal
and that corrections to the Nambu-Goto action may appear only at very high order in $1/\sigma_0 R^2$, 
thus it is well possible that the most important source of corrections is not due to higher order non-universal terms in the effective string action but
to the coupling to the massive modes. Indeed, corrections to the excited string states in (3+1) dimensions were recently observed in~\cite{Athenodorou:2010cs} and
associated to a possible coupling to massive modes\footnote{Notice however that no deviation was observed in a similar calculation in (2+1) 
dimensions~\cite{Athenodorou:2011rx}.}. Moreover it was recently observed that in the 3d Ising gauge model corrections to the effective string potential larger than those
predicted by the universality theorems are present both in the torus (interface)~\cite{Caselle:2007yc} 
and in the cylinder (Polyakov loop correlators)~\cite{Caselle:2010pf} geometries. All these examples suggest that
a more detailed study of these intrinsic massive corrections would be  important non only from a theoretical point of view but also 
to improve our comparisons with numerical simulations.

However, performing this analysis in the original (d+1) dimensional LGT turns out to be very demanding from a numerical point of view. 
Thus  in this paper we decided to adopt an alternative route. 
Following the approach that we recently developed in \cite{cdgjm05,Caselle:2006wr}, we studied the intrinsic width
in the vicinity of the deconfinement transition (but still in
the confining phase)  looking at 
suitable correlators in a d dimensional spin model which, in the framework the well known Svetitsky-Yaffe 
analysis~\cite{sy82}, is known to 
represent an accurate effective description of the original  LGT  in the vicinity of the deconfinement transition.

\section{Three proposals for the evaluation of the intrinsic width in the finite temperature LGTs} 

Finding a lattice observable to measure the the intrinsic width of the flux tube at finite temperature  
is a rather nontrivial task.

At low temperature the simplest and most natural option would be to identify the intrinsic 
width with the constant term which appears in the function which describes the dependence of the flux tube thickness on the interquark
distance $R$.
Following~\cite{lmw80} we expect at low temperature the following behaviour for the effective width (in adimensional units)
\eq
w^2=\frac{d-2}{2\pi\sigma_0}\log(R/R_c)
\label{f1}
\en
where we denote with $\sigma_0$ the zero temperature string tension of the gauge model.
This prediction  was confirmed by high precision numerical simulations both in the 3d gauge Ising model~\cite{cgmv95} model and in the 3d 
SU(2) Yang Mills theory~\cite{Gliozzi:2010zv}.

The scale $R_c$, measured in units of $1/\sqrt{\sigma_0}$ turned out to be almost the same in the two models: $R_c\sqrt{\sigma_0}=0.337(18)$ in Ising  and
$R_c\sqrt{\sigma_0}=0.364(3)$ for SU(2).

 Eq.(\ref{f1}) can be rewritten in adimensional units as
\eq
\sigma_0 w^2=\frac{d-2}{2\pi}\log(R\sqrt{\sigma_0}) +b
\label{f2}
\en
The fit of the numerical data to this law gives very good $\chi^2$ values down to values of the $R\sim 1/\sqrt{\sigma_0}$. 
Deviations below this threshold may be assumed to be due to the intrinsic string thickness $\xi_I$ that we are looking for. 
In this way we would obtain both for Ising and for SU(2)
 $\xi_I\sqrt{\sigma_0}\sim\sqrt{b}\sim 0.4-0.5$. The major drawback of this qualitative estimate is that the value of $b$ rather strongly  
depends on the assumptions on
the shape of the flux tube. This is not a gaussian (see for instance \cite{Muller:2004vv,Kopf:2008hr} for a detailed study in the 3d Ising case) 
and there is not a common consensus on which should be the functional form. 
Indeed, as we shall see below the two issues of a non zero intrinsic
width and non-gaussian shape of the flux tube are deeply interconnected.

The situation becomes even worse if one is interested in a finite temperature estimate of the intrinsic width since in this regime 
the square width of the flux tube increases linearly instead of logarithmically~\cite{Allais:2008bk} and the constant term gets contributions 
also from the Nambu-Goto effective string action thus making it impossible to disentangle the intrinsic width.

It is thus important to find an alternative observable which could allow a direct estimate of $\xi_I$.

Three possible candidates are:

\begin{description}
\item{a]}
The large distance transverse behaviour of the plaquette - Polyakov loop correlator expectation value.

\item{b]}
The second moment in the transverse direction of the completely connected  four point function of two Polyakov loops and two plaquettes~\cite{gliozzi}. 

\item{c]}
The large distance transverse behaviour of the  completely connected  four point function of two Polyakov loops and two plaquettes. 

\end{description}

In order to characterize these proposals let us first review the standard way in which the effective width of the flux tube
is evaluated  on the lattice.
In a finite temperature setting the lattice operator which is used to evaluate the flux through a plaquette $p$ of the lattice is:
\eq 
\bra\phi(p;P,P')\ket=\frac{\bra P P'^\dagger~U_p\ket}{\bra PP'^\dagger \ket}-\bra U_p\ket 
\label{flux2} 
\en 
where $P$, $P'$ are two Polyakov loops separated by $R$ lattice spacings and $U_p$ is 
the operator associated with the plaquette $p$. Let us choose $p$ to be  equidistant from the two Polyakov loops (i.e. in the space 
perpendicular to the plane defined by the two Polyakov loops and intersecting this
plane exactly at $R/2$). 
Then the width of the flux tube $w$ is defined as:
\eq
w^2(R,L)=\frac{\sum_{\vec h} \vec h^2 \bra\phi(\vec h;R,N_t)\ket}{\sum_{\vec h} \bra\phi(\vec h;R,N_t)\ket}
\label{w1}
\en
where $R$ denotes the distance between the two Polyakov loops $N_t$ the extension of the lattice in the compactified time direction 
(i.e. the inverse temperature: $T\equiv  1/N_t$) and $\vec h$ the displacement of $p$ from the $P$ $P'$ plane.

With this definition  the first proposal mentioned above corresponds to studying the $|h|>>R$ limit of $\bra\phi(\vec h;R,N_t)\ket$. If in this limit the 
correlator shows an exponentially decaying behaviour then we may estimate $\xi_I$ as 
$\bra\phi(\vec h;R,N_t)\ket \sim \exp(-|h|/\xi_I)$. The rationale behind this proposal is that in this limit the effective width, 
(which in the vicinity of the deconfinement transition is proportional to $\sqrt{R}$) becomes negligible 
and the scale which drives the exponential decay must be the intrinsic width. 
This proposal is clearly inspired by the definition of the  
London penetration length (see for instance \cite{Cardaci:2010tb} and references
therein) and requires, to be defined, that the flux tube should not have a gaussian shape (as instead predicted by the Nambu-Goto action). This is another way to say that
we are looking to effects beyond the Nambu-Goto effective string description. 

The second proposal above corresponds to looking at the four point function (see fig.\ref{fig})
\eq 
{\bra\phi_I(p,p';P,P')\ket=\bra P P'^\dagger~U_p~U_{p}'\ket}_c 
\label{flux3} 
\en 
where ${\bra~~~\ket}_c$ denotes the completely connected expectation value and, as above, the two Polyakov loops are located
at a distance $R$ 
while the two plaquettes are located at the
opposite sides with respect to the $P$ $P'$ plane at a distance $y$ from the plane 
(We refer to fig.\ref{fig} for the geometrical setting. Notice, to avoid confusion that in fig.\ref{fig} we set $r=R/2$). 
This quantity can be considered, in analogy to the flux $\phi$ defined in eq.(\ref{flux2}), as a sort of "intrinsic flux" $\phi_I$. In fact this quantity becomes a $\delta$
function centered in $y=0$ if we
try to evaluate it in a stringy framework (and in particular assuming the Nambu-Goto action in the physical gauge) 
in which any expectation value is obtained as a sum over single valued surfaces which cannot simultaneously pass through $P$ and $P'$ unless $P=P'$.

Thus the second moment of $\phi_I$ is likely to be an estimator of the intrinsic width (in exact analogy with the use of the second moment of $\phi$ as an estimator for the
effective width $w$). 

Finally, the peculiar properties of the quantity $\phi_I$ suggest another interesting possibility to estimate $\xi_I$. 
In fact, it is conceivable to expect that the intrinsic width should
be related to the change in the value of the correlation length 
between two plaquettes due to the presence of a flux tube (mimicked by the correlator of the two Polyakov loops
between them). The simplest way to measures this quantity is to study the ratio 
\eq 
\frac{\bra P P'^\dagger~U_p~U_{p}'\ket_c}{\bra U_p~U_{p}'\ket_c}
\label{flux3bis} 
\en 

Then, according to the above conjecture, for any value of $R$ this ratio
is expected to decay in the large $y/R$ limit 
as $\sim \exp(-|y|/\xi_I)$ where $\xi_I$ is the intrinsic width.

The main problem is that these observables are rather difficult to estimate numerically. For this reason, following the approach that
we developed in our previous papers
\cite{cdgjm05,Caselle:2006wr}, we decided to address first the problem in the framework of the Svetitsky-Yaffe conjecture~\cite{sy82} looking at 
suitable correlators in 2d spin models. 

More
precisely we chose to study gauge models in (2+1) dimensions with a second order deconfinement transitions and with a gauge group whose center is $Z_2$. The two simplest
cases belonging to this class are the (2+1) dimensional gauge Ising model and the (2+1) SU(2) Yang-Mills theory. These models in the vicinity of the deconfinement transition belong to the
same universality class of the 2d Ising model. Moreover the confining phase below the transition corresponds in this mapping to the energy perturbation of the 2d Ising
universality class.  
 In the following section we shall construct a dimensionally reduced
projection of the above correlators. We shall first review the general formalism and recall how the standard 
effective width may be obtained in
this framework. Then we shall address the three proposal mentioned above to evaluate $\xi_I$.
Taking advantage of the integrability of the energy perturbation of the 2d Ising field theory, we shall be able to evaluate exactly the large distance behaviour of the correlators involved in these observables, and to eventually extract the value of the intrinsic width. We shall show that, as
expected, the intrinsic width does not depend on the interquark distance and that the three different ways that we propose to evaluate
it consistently give the same answer.

\section{Dimensional reduction and the Svetitsky Yaffe conjecture.} 
According to the Svetitsky--Yaffe conjecture~\cite{sy82}, if we choose to study a (d+1) LGT (with gauge group $G$)
with a second order deconfinement phase transition and if the gauge group $G$ is such that the $d$ dimensional spin model with (global) 
symmetry group the center of $G$ also has a continuous symmetry breaking phase transition then
the two critical points must  
belong to the same universality class and we can use the spin model as an effective theory description for the  
(d+1) dimensional LGT in the neighbourhood of the deconfinement transition. 
This is the case if we choose for instance the (2+1) $SU(2)$ LGT  or the (2+1) Ising gauge model. Both have a
second order deconfinement phase transition and in both cases one can use the 2d spin Ising model as an effective theory 
in the neighbourhood of the deconfinement point.

In this effective description the Polyakov loops of the LGT are mapped into the spins of the Ising  
model, the confining phase of the LGT into the high temperature phase of the spin model and the plaquette operator of the 
LGT is mapped into the energy operator of the spin Ising model.
The reason why such an approach is particularly effective in the present setting, is the integrability of the Ising field theory in zero magnetic field. This property allows for the calculation of multipoint correlation functions in the regime of large distances via the spectral expansion over form factors. The latter can be calculated exactly, and used as building blocks to write down analytic expressions for the large distance behavior of various correlators. In the following we will be interested in the asymptotic behaviour of three- and four-point correlators involving spin and energy operators. We refer the reader to \cite{Yurov, Karowski, Smirnov,cdgjm05,Caselle:2006wr} for technical details on the calculations.    

It is important to stress that this approach allows only to estimate observables in the (2+1) LGT 
which diverge as the deconfinement point is approached. Thus an implicit assumption
behind our study is that also $\xi_I$ should be a quantity of this type, i.e. it should diverge in the deconfinement limit. 
We shall see below that this is indeed the case, a results which represents a non trivial 
self-consistency check of our approach.

\subsection{The 3-point correlator $\langle \sigma \epsilon \sigma \rangle$: the first proposal to estimate the intrinsic width.} 

Let us recall the basic operator mapping between spin model and LGT. The Polyakov loop $P$, being the order parameter of the transition, is mapped onto the spin operator $\sigma(x)$. The operator of the spin model corresponding to the plaquette operator $U$ was shown to be the energy density $\epsilon(x)$ in \cite{plaquette} (see~\cite{Caselle:2006wr} for a detailed
discussion of this mapping).  \\
Following the above discussion 
the operator which measures the flux density in presence of the Polyakov loops pair in the LGT is mapped into the three points function $\langle \sigma \sigma  \epsilon \rangle$.
In the particular case of the 2d Ising model this correlator can be 
evaluated in the vicinity of the critical point using the spectral expansion, and
 (see next section for further details and 
\cite{Yurov} for a review)
leading to the following  expression for
the "flux"  distribution~\cite{Caselle:2006wr}
\eq 
S(R,y) \ = \bra \sigma \sigma \epsilon\ket \ \sim \frac{ (F_1^\sigma)^2  2 R}{4y^2+R^2} \, {e^{-m \sqrt{4y^2+R^2}}}.  
\label{nongaussian} 
\en 
 where $y$ denotes the transverse direction,  $m$ is the mass of the 2d Ising model
and a large $mR$ limit is assumed.

The effective width of the flux tube is then given by the ratio

\eq 
w^2(R) \ = \frac{\int_{-\infty}^{\infty} dy \, y^2 \, S(R,y)}{\int_{-\infty}^{\infty} dy \, S(R,y)}  
\label{sy1}
\en 

This ratio is easy to evaluate in the in the large $mR$ limit 
(see~\cite{Caselle:2006wr,Caselle:2010zs} (Notice, to avoid confusion, 
that in \cite{Caselle:2006wr} we used the variable $r\equiv R/2$ and that we 
evaluated the unnormalized width, i.e. only the numerator of eq.(\ref{sy1}))) 

leading to the following result:
\eq 
w^2(R) \simeq \  \frac{1}{4} \, \frac{R}{m} + \dots.
\label{sy2}
\en 
where the dots stay for terms constant or proportional to negative powers of $R$.

The intrinsic width in which we are interested can be obtained looking at the large $y$ limit of
 eq.(\ref{nongaussian}). In this limit we obtain
 
\eq 
S(R,y) \ \sim \ \frac{ (F_1^\sigma)^2  2 R}{4y^2} \, {e^{-2 m y}}.  
\label{largey} 
\en 

from which we find $\xi_I=1/2m$. As mentioned in the previous section the reason for which we could find a non trivial result for $\xi_I$
is related to the fact that $S(R,y)$ has not a purely gaussian behaviour in the $y$ variable.

\subsection{The 4-point correlator $\langle \epsilon \sigma \epsilon \sigma \rangle$: the second proposal to estimate the intrinsic width} 

We want to study the 4-point correlation function of the form 
\bea 
\langle \epsilon(x_1) \sigma(x_2) \epsilon(x_3) \sigma(x_4)\rangle
\eea 
in the high temperature phase of the 2d Ising model in zero magnetic field.  Since we are interested in its large distance behaviour, we may use also in this case  the Form Factors approach.
 
\begin{figure} 
\centering 
\includegraphics[height=7cm]{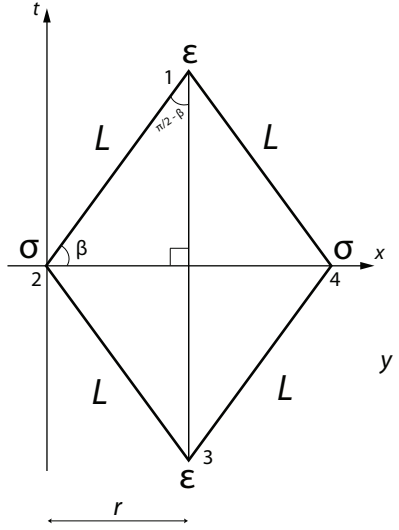} 
\vskip0.5cm 
\caption{Rhombus configuration for the four-point function $\langle \epsilon\sigma \epsilon \sigma \rangle_c^\Diamond $, eq.~(\ref{tetrarhombus}).} 
\label{fig} 
\end{figure}

To evaluate the intrinsic width we are interested in particular in 
the connected correlation function for a generic rhombus of side $L$, as described in figure \ref{fig}. 
In such a particular case, it is enough to consider the first two contributions to the spectral expansion  (we proceed in close analogy with the analysis of \cite{cdgjm05} and \cite{Caselle:2006wr}), 
\bea 
&& \langle \epsilon \sigma \epsilon \sigma \rangle_c^\Diamond  =  
\int_{-\infty}^{\infty} \frac{d \theta_1 \dots d \theta_3}{ (2 \pi)^3} \, (F_1^\sigma) \, F_2^\epsilon  
(\theta_{12}- i \beta + 2 i \pi) \,  F_2^\epsilon  
(\theta_{23}+ i \beta ) e^{- m L(\cosh \theta_1+\cosh \theta_2 )-2 m y \cosh \theta_3} + \nonumber \\ 
&&+\int_{-\infty}^{\infty} \frac{d \theta_1 \dots d \theta_4}{2!~ (2 \pi)^4} \, (F_1^\sigma) \, F_2^\epsilon  
(\theta_{12}) \,  F_2^\epsilon  
(\theta_{34} - 2 i \beta+ i \pi)  \cdot [ F_3^\sigma(\theta_{31}+ 2 i \beta, \theta_{32}+2 i \beta , \theta_{12}) ] \, \nonumber \\ 
&& \cdot e^{- m L(\cosh \theta_1+ \dots + \cosh \theta_4) } + \dots 
\label{tetracorr} 
\eea 
where we listed the leading and next-to-leading contributions to the spectral expansion.
Inserting the explicit expressions for the form factors, we have
\bea 
&&\langle \epsilon \sigma \epsilon \sigma \rangle_c^\Diamond  = 
-\frac{ ( m F_1^\sigma )^2}{4 \pi} 
\int_{-\infty}^{\infty} d \theta_1 \dots d \theta_3 \,
 \sinh (\theta_{12}/2-i \beta) \,  \sinh (1/2(\theta_{23} + i \beta )) \cdot
 \nonumber \\ 
&& e^{- m L(\cosh \theta_1+\cosh \theta_3 )-2 m y \cosh \theta_2} + \nonumber \\ 
&&-\frac{ (2 \pi m F_1^\sigma )^2}{2!~ (2 \pi)^4} 
\int_{-\infty}^{\infty} d \theta_1 \dots d \theta_4 \, \sinh (\theta_{12}/2) \,  \cosh (1/2(\theta_{34} - 2i \beta )) \cdot \nonumber \\ 
&& \cdot \tanh (\theta_{12}/2) \,  \tanh (1/2 (\theta_{13}+2 i \beta)) \, \tanh (1/2 (\theta_{23}+2 i \beta))
 e^{- m L(\cosh \theta_1+ \dots + \cosh \theta_4)  } + \dots
\eea 
Using the following property in the second term
\bea
&& \tanh (\theta_{12}/2) \,  \tanh (1/2 (\theta_{13}+2 i \beta)) \, \tanh (1/2 (\theta_{23}+2 i \beta)) = \nonumber \\
&&  = \tanh (\theta_{12}/2) -  \tanh (1/2 (\theta_{13}+2 i \beta)) +  \tanh (1/2 (\theta_{23}+2 i \beta)) 
\eea
and after some long but straightforward calculations we can reduce the correlator to this form
\bea
&&\langle \epsilon \sigma \epsilon \sigma \rangle_c^\Diamond  =
(m F_1^\sigma )^2 \frac{e^{-2 mL}}{mL} 
\big[ K_0(2 m y) - \sin \beta \, K_1(2 m y) \big] +  \nonumber \\
&&+\frac{ ( m F_1^\sigma )^2 e^{-2 mL}}{2!~ mL } 
\int_{-\infty}^{\infty} d \theta \, \big[ p_\beta (\theta) K_0(2 m L \cosh (\theta/2)) + q_\beta (\theta) K_1(2 m L \cosh (\theta/2)) \big] + \dots
\nonumber \\
&&
\label{tetrarhombus}
\eea
where
\bea
&& p_\beta (\theta) = \cos \beta \, \frac{\sinh^2 \theta/2}{\cosh \theta/2} -
\frac{\sinh^2 \theta/2-\sin^2 \beta}{\cosh \theta/2 \, \cos \beta(1+\tanh^2 \theta/2 \tan^2 \beta)} \nonumber \\
&&  q_\beta (\theta) = -  \frac{\sin^2 \beta}{\cos \beta} \,
\frac{1}{\cosh^2+ \sinh^2 \theta/2 \tan^2 \beta}
\eea
and
\bea
L= \sqrt{r^2 + y^2}; \ \ \ \ \ \beta = \arcsin \frac{y}{\sqrt{r^2 + y^2}}.
\eea

Paralleling the usual treatment for the flux tube thickness, we can compute the quantity 
\bea
w^2_I = \frac{\int_{-\infty}^{\infty} dy \, y^2 \langle \epsilon \sigma \epsilon \sigma \rangle_c^\Diamond }{\int_{-\infty}^{\infty} dy \langle \epsilon \sigma \epsilon \sigma \rangle_c^\Diamond}
\eea
using the leading, large-$r$ behaviour of $\langle \epsilon \sigma \epsilon \sigma \rangle_c^\Diamond$. In the limit $r/y \to \infty$, we have
\bea
\int_{-\infty}^{\infty} dy \, y^2 \langle \epsilon \sigma \epsilon \sigma \rangle_c^\Diamond = (m F_1^\sigma )^2 \left( \frac{2 r^2}{m} e^{-2 m r} \, \frac{\pi}{16(mr)^3} +
\dots \right)
\eea
\bea 
\int_{-\infty}^{\infty} dy \,  \langle \epsilon \sigma \epsilon \sigma \rangle_c^\Diamond  = (m F_1^\sigma )^2 \left( \frac{2 }{m} e^{-2 m r} \, \frac{\pi}{4 \, mr} + \dots
\right)
\eea
Substituting in the formula for $w^2_I$, we obtain
\bea
w^2_I = \frac{\int_{-\infty}^{\infty} dy \, y^2 \langle \epsilon \sigma \epsilon \sigma \rangle_c^\Diamond }{\int_{-\infty}^{\infty} dy \langle \epsilon \sigma \epsilon \sigma \rangle_c^\Diamond} = \frac{1}{4 m^2} + \dots 
\ \ \ \ \ \ 
\longrightarrow \ \ \ \ \ \ \xi_I = \sqrt{w^2_I} =  \frac{1}{2 m} 
\eea
in full agreement with the result of the previous sections.

\subsection{Large-$y$ behaviour: the third proposal to estimate the intrinsic width.} 

Following the third proposal discussed above, we are interested in computing the leading behaviour of the ratio
 $\langle \epsilon\sigma \epsilon \sigma \rangle_c^\Diamond/ \langle \epsilon (y) \epsilon(0) \rangle_c$ in the limit of large transverse coordinate $y$.
  Introducing the variable $x = r/y$, and taking the small $x$ limit we have
\bea
\frac{\langle \epsilon \sigma \epsilon \sigma \rangle_c^\Diamond}{\langle \epsilon (y) \epsilon(0) \rangle_c}  =
 (F_1^\sigma) ^2 ~ \Bigg[ \frac{1}{2 \sqrt \pi (my)^{1/2}}+\dots+\frac{1}{mr} \Bigg( \frac{1}{2}+\dots \Bigg) \Bigg] e^{-2 my} + \dots
\eea
where
\bea
\langle \epsilon (y) \epsilon(0) \rangle_c  = 
m^2 \, \left[ K_1^2(my) - K_0^2(my) \right] = \frac{ \pi m^2}{2 \, (my)^2} e^{-2 my} + \dots 
\eea
from which we can read off that the exponential decay is ruled by $\xi_I=(2m)^{-1}$, which can be interpreted as the intrinsic width. 

\section{Discussion.}

\subsection{Temperature dependence of the intrinsic width} 

We have seen in the previous section that the three approaches consistently give the same result $\xi_I=1/(2m)$. 
It is interesting to convert this result in the language of the original
(2+1) dimensional LGT.

Looking at the behaviour of the correlator of two Polyakov loops and comparing it with its spin model projection we immediately identify 
(see for instance \cite{Caselle:2010zs})

\eq 
  m=\sigma(T)/T
\en 
and hence 
\eq 
  \xi_I=\frac{T}{2\sigma(T)}
\label{final}
\en 

where $T=1/N_t$ is the temperature and $\sigma(T)$ is  the temperature dependent string tension which vanishes at $T=T_c$ and has a critical behaviour 
given by 
\eq 
  \sigma(T)=\sigma_0  
  \left(1-\frac{T}{T_c} 
  \right)^{\nu}. 
\label{sigmatsec}
\en
where $\nu$ is (following the Svetitsky-Yaffe conjecture) the
critical index of the 2d spin model (i.e. $\nu=1$)
and $\sigma_0$ is the usual string tension (which in this framework is the $T\to 0$ limit of $\sigma(T)$). Thus our analysis shows that in the vicinity 
of the deconfinement
transition the intrinsic width depends on the temperature and (as the effective width) diverges at the deconfinement transition. As mentioned above this is an important
self-consistency check of our whole approach.

Eq(\ref{final}) is the main result of our paper and in the vicinity of the deconfinement point, it may be rewritten as:
\eq
\xi_I(T)=\frac{T}{2\sigma_0}\left(1-\frac{T}{T_c} 
  \right)^{-1}
\en

It is interesting to observe that, using the well known relationship between $T_c$ and $\sigma_0$ proposed by Olesen in the framework of the Nambu-Goto effective string
model
\cite{Olesen:1985ej}, which in (2+1) dimensions reads $T_c^2=3 \sigma_0/\pi$ we obtain in the vicinity of the deconfinement point

\eq
\xi_I(T)\sqrt{\sigma_0}\ \sim \frac{3}{2\sqrt{\pi}}\left(1-\frac{T}{T_c} 
  \right)^{-1}
\en

which looks like a finite T extension of the result $\xi_I(T=0)\sqrt{\sigma_0}\sim 0.4-0.5$ quoted above for the zero temperature limit of the intrinsic width 

\subsection{Range of validity of the Effective String description} 

It is intersting to study the implications of our results for the range of validity of the effective string decription. As we mentioned in the introduction the 
effective string description is expected to hold down to scales of the order of
$R\sim 1/\sqrt{\sigma(T)}$. 
However we certainly do not expect it to hold below $\xi_I$ and we see from  eq.(\ref{final}) that as the deconfinement temperature is approached
$\xi_I$ becomes indeed the limiting scale since it diverges as $1/\sigma(T)$. This should be carefully taken 
into account when comparing numerical simulations with effective
string predictions in the vicinity of the deconfinement point and suggests that this regime could be the optimal one to observe signatures of this intrinsic scale.

\subsection{Non-analytic terms in the spin-spin correlator} 

We mentioned in the introduction that the massive intrinsic excitations of the flux tube should manifest themselves not only in the
intrinsic width but also as massive corrections to the interquark potential. It is not obvious that 
we can trust our approach also to the level of subleading corrections, however it is 
intriguing to observe that if we perform a large distance expansion of the
spin-spin correlator (which is the 2d analogue of the Polyakov loops correlator) we find:

\bea
\langle \sigma(r) \sigma(0) \rangle = \frac{(F_1^{\sigma})^2}{\sqrt{2 \pi}} 
\frac{e^{- mr}}{(mr)^{1/2}} ~ \Bigg[ 1+ \frac{1}{64 \pi} \frac{e^{-2 mr}}{(mr)^4} \dots \Bigg]
\eea
from which, assuming the validity of the dimensional reduction approach also at the level of subleading corrections, we can extract
the non-analytic contribution to the interquark potential ($T$ is the temperature in the gauge theory)
\bea
\frac{V(r,T)}{T} = - \log \langle \sigma(r) \sigma(0) \rangle = 
mr + \frac{1}{2} \log mr - \frac{1}{64 \pi}  \frac{e^{-2 mr}}{(mr)^4} + \dots.
\eea

Looking at the exponential term in the correlator we may easily extract the mass of this intrinsic excitation 
which turns out to nicely 
agree with the previous estimates, $\xi_I = (2m)^{-1}$. 
It is interesting to notice that a similar result was recently discussed in \cite{Vyas:2010wg} in the framework of a holographic
model for the intrinsic width.

\vskip1.0cm {\bf Acknowledgements.}

We warmly thank  F.~Gliozzi for many useful discussions and suggestions. M.C. thanks all the participants of the {\it Confining Flux Tubes and Strings} 
Workshop at ECT$^\star$, Trento during July 2010 for several useful discussions, which partially stimulated the present work.

\end{document}